\documentclass[aps,prapplied,reprint,showpacs,amsmath,amssymb,floatfix,superscriptaddress,noeprint]{revtex4-1}
\usepackage{color}
\usepackage[colorlinks,linkcolor=blue,urlcolor=blue,citecolor=blue,anchorcolor=blue]{hyperref}
\usepackage{graphicx}
\usepackage[capitalise]{cleveref}
\usepackage{physics}
\usepackage{siunitx}
\usepackage{soul}

\newcommand{\nth}[1]{N^{\mathrm{th}}_{#1}}

\begin{document}
\title{Demonstration of Quantum Advantage in Microwave Quantum Radar}
\date{\today}

\author{R. Assouly}
\affiliation{Ecole Normale Sup\'erieure de Lyon,  CNRS, Laboratoire de Physique, F-69342 Lyon, France}
\author{R. Dassonneville}
\affiliation{Ecole Normale Sup\'erieure de Lyon,  CNRS, Laboratoire de Physique, F-69342 Lyon, France}
\author{T. Peronnin}
\affiliation{Ecole Normale Sup\'erieure de Lyon,  CNRS, Laboratoire de Physique, F-69342 Lyon, France}
\author{A. Bienfait}
\affiliation{Ecole Normale Sup\'erieure de Lyon,  CNRS, Laboratoire de Physique, F-69342 Lyon, France}
\author{B. Huard}
\email{benjamin.huard@ens-lyon.fr}
\affiliation{Ecole Normale Sup\'erieure de Lyon,  CNRS, Laboratoire de Physique, F-69342 Lyon, France}

\maketitle

\textbf{While quantum entanglement can enhance the performance of several technologies such as computing, sensing and cryptography,
its widespread use is hindered by its sensitivity to noise and losses.  Interestingly, even when entanglement has been destroyed~\cite{Knill1998, Datta2008,Lloyd2008}, some tasks still exhibit a quantum advantage $Q$, defined by a $Q$-time speedup, over any classical strategies.
A prominent example is the quantum radar~\cite{Lloyd2008}, which enhances the detection of the presence of a target in noisy surroundings. To beat all classical strategies, Lloyd~\cite{Lloyd2008} proposed to use a probe initially entangled with an idler that can be recombined and measured with the reflected probe. 
Observing any quantum advantage requires exploiting the quantum correlations between the probe and the idler. It involves their joint measurement~\cite{Bradshaw2017} 
or at least adapting the idler detection to the outcome of the probe measurement~\cite{Shi2022}.
In addition to successful demonstrations of such quantum illumination protocols at optical frequencies~\cite{Zhang2015,Xu2021}, the proposal of a microwave radar~\cite{Barzanjeh2015,Pirandola2018}, closer to conventional radars, gathered a lot of interest. 
However, previous microwave implementations~\cite{Bourassa2020,Luong2018,Luong2020,Chang2019,Barzanjeh2020,Livreri2021,Hosseiny2022} have not demonstrated any quantum advantage as probe and idler were always measured independently~\cite{Shapiro2020,Jonsson2020,Sorelli2020}. 
In this work, we implement a joint measurement using a superconducting circuit and demonstrate a quantum advantage $Q>1$ for microwave radar.
Storing the idler mitigates the detrimental impact of microwave loss on the quantum advantage, and the purity of the initial entangled state emerges as the next limit~\cite{DiCandia2021}. 
While the experiment is a proof-of-principle performed inside a dilution refrigerator,
it exhibits some of the inherent difficulties in implementing quantum radars such as the limited range of parameters where a quantum advantage can be observed or the requirement for very low probe and idler temperatures.}

\begin{figure}[h!]
    \centering
    \includegraphics[width=85mm]{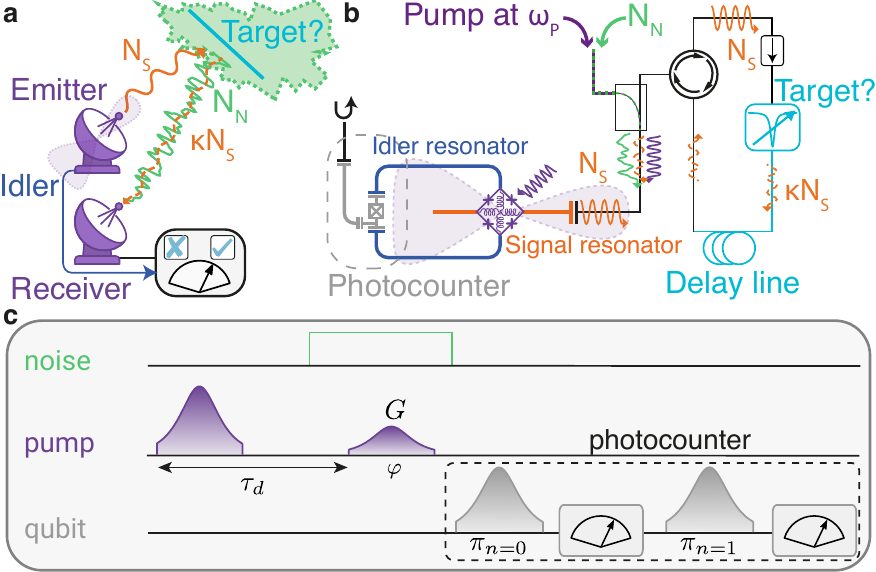}
    \caption{\label{fig:figure_1}Quantum radar principle and implementation. \textbf{a}) An emitter sends a probe signal with photon number $N_S$ to determine the presence of a target using the least possible number of attempts. The signal is reflected or not from the target, with reflectivity $\kappa$, in a thermal environment with mean photon number $N_N$. A receiver processes all reflected signals and decides whether the target is present or not. Quantum probes can be initially entangled with an idler whose processing by the receiver leads to faster determination and thus a quantum advantage compared to any classical probe. The advantage crucially relies on the exploitation of quantum correlations between idler and received signal.
    \textbf{b}) Superconducting circuit (left) probing a target composed of a delay line and tunable notch filter (right). It comprises a non-linear device (purple) generating and decoding entangled pairs between signal mode (orange) and idler mode (blue). A transmon qubit (grey) completes the joint measurement. The entangling pump and thermal noise background are injected through a directional coupler into the signal resonator port. \textbf{c}) Pulse sequence of the quantum radar experiment. The phase difference $\varphi$ and delay $\tau_d$ between pump pulses, as well as the gain $G$ of the second pump pulse can all be tuned. The dashed box represents the measurement by the qubit of the effective photon number $\nu$ in the idler resonator for quantum radar but can be replaced by other photocounting schemes for calibration purposes (see Methods).}

\end{figure}

We focus on the simplest radar protocol, where the goal is to detect whether a target is present with a minimum number $M$ of attempts. Each attempt corresponds to using a single microwave mode in time-frequency space to probe the target, with the constraint that the probe contains a fixed number $N_S$ of signal photons on average (Fig~\ref{fig:figure_1}a) and is detected in a noise background of $N_N$ photons. We consider that all other parameters are known: target position, speed and reflectivity $\kappa$. Several metrics can quantify the performance of a radar. 
We choose the error exponent defined as $\mathcal{E}= \lim_{M \to \infty} -\frac{1}{M}\log P_\mathrm{error}(M)$, which means that the error probability  $P_{\mathrm{error}}(M)$ is logarithmically equivalent to $e^{-\mathcal{E}M}$. 
For simplicity, we assume no prior knowledge on the target state: initially the target is present with a probability $\frac{1}{2}$. Under the assumptions of the central limit theorem, the number of required attempts to reach a given error probability scales as $1/\mathcal{E}$. The quantum advantage can thus be defined as $Q=\mathcal{E}/\mathcal{E}_\mathrm{cl}$, where $\mathcal{E_\mathrm{cl}}$ is the error exponent of the best classical strategy. 

Given a certain probe state, the largest achievable error exponent for any measurement apparatus is the so-called quantum Chernoff bound~\cite{Audenaert2007}.
De Palma and Boregaard~\cite{DePalma2018} showed that the best classical strategy (i.e. without quantum memory) is to use a coherent state as a probe, which gives an optimum $\mathcal{E}_\mathrm{cl} = \frac{\kappa N_S}{4 N_N}$. This limit is asymptotically reached by a homodyne measurement in the large noise ($N_N \gg 1$) limit~\cite{Guha2009}.
Quantum strategies rely on initially entangling the probe with an idler~\cite{Lloyd2008}. The quantum Chernoff bound for quantum radar is $\mathcal{E}_{\mathrm{max}} = \frac{\kappa N_S}{N_N}$ in the low signal $N_S \ll 1$, high noise $N_N \gg 1$ regime~\cite{DePalma2018}, which shows that the quantum advantage is at best $Q_\mathrm{max}=4$ for radars. 
Effectively, it can be reached using one mode of a two-mode squeezed vacuum state (TMSV) to illuminate the target~\cite{Tan2008,Nair2020}. However, there is no known detector that can reach this advantage $Q_\mathrm{max}=4$ without a global joint measurements of $M$ modes of all attempts~\cite{Zhuang2017, Shi2022}. Using simpler pairwise joint measurements instead~\cite{Guha2009,Calsamiglia2010,Sanz2017}, it is nevertheless possible to reach $Q=2$ with $\mathcal{E}_{\mathrm{pair}} = \frac{\kappa N_S}{2N_N}$.

Here we implement pairwise joint measurements using a superconducting circuit~\cite{Peronnin2019, dassonneville2020, Dassonneville2021} that also generates the TMSV states~\cite{Eichler2011,Wilson2104,Flurin2012,Menzel2012a}, and stores the idler mode while the signal probe travels. We then experimentally determine the error exponent of this quantum radar for various signal and noise photon numbers. To ensure a fair determination of the experimental quantum advantage $Q$, the absolute best classical error exponent $\mathcal{E_{\mathrm{cl}}}$ must be determined. Previous microwave radar experiments managed to exceed the error exponent of one instance of classical radar~\cite{Bourassa2020,Luong2018,Luong2020,Chang2019,Barzanjeh2020,Livreri2021,Hosseiny2022}, but could not break the  classical upper bound $\mathcal{E_{\mathrm{cl}}}$.  A central challenge of the experiment thus consists in performing precise calibrations of the target and radar parameters $\kappa$, $N_S$ and $N_N$.

\paragraph*{Microwave Quantum Radar Implementation}
Our superconducting device contains two resonators: a signal resonator whose lifetime is set by its coupling to a transmission line and a much longer lived idler resonator. The circuit is operated at \SI{15}{mK} (Fig.~\ref{fig:figure_1}b). The signal resonator, which emits and receives the probe signal, has frequency $\omega_S/(2\pi) = \SI{10.20}{GHz}$  and is coupled to a transmission line at a rate $\gamma_S/(2\pi)=\SI{25}{MHz}$. The idler resonator has frequency $\omega_I/(2\pi) = \SI{3.74617}{GHz}$ and a decay rate of $\gamma_I/(2\pi) = \SI{40}{kHz}$. 
The two resonators are coupled by a Josephson Ring Modulator (JRM, purple in Fig.~\ref{fig:figure_1}b)~\cite{Bergeal2010, Bergeal2010b,Roch2012}. 

We start each of the $M$ detection attempts by first applying a pump tone at a frequency $\omega_P = \omega_S + \omega_I$ for \SI{28}{ns}. This tone generates a TMSV state between the idler and the signal modes. The latter quickly exits its resonator~\cite{Flurin2015}, propagates to the target and, when the target is present, comes back attenuated by a factor $\kappa$ (see Fig.~\ref{fig:figure_1}b-c). The target is composed of a circulator and a flux tunable notch filter (see Methods) followed by a \SI{12}{m}-long coaxial cable, allowing us to tune the target reflectivity in situ from $\kappa$ (target is present) to a value two orders of magnitude lower (target is absent). The reflected probe is then combined with thermal noise injected via a weakly coupled auxiliary line. 
The noise is generated at room temperature by amplifying the Johnson-Nyquist noise of a \SI{50}{\ohm} resistor with a tunable gain (see Methods).

A quantum advantage can only be observed for $N_N>1$, yet the generation of the TMSV state requires the signal resonator to be as cold as possible. We thus only switch on the noise source after the TMSV state has been prepared. 

The joint measurement is finally performed as follows. We drive the JRM with a pump at $\omega_P$ with the same nominal phase as the first pulse. The resulting two-mode squeezing operation recombines the  signal and idler in such a way that the final idler state encodes the presence or absence of quantum correlations between reflected signal and stored idler~\cite{Guha2009}. The proposal of Guha and Erkmen~\cite{Guha2009} recommends measuring the final number of photons in the idler mode to reach up to $Q=2$.

Despite the large thermal background of the reflected signal, the number of photons $\nu$ in the idler remains low  after the recombination (see Methods). Thus, measuring whether there is zero photon or more is in theory sufficient to observe a quantum advantage. Instead, we truncate the photon counting to $2$ to enhance the quantum advantage as much as possible. The measurement starts by conditionally exciting the qubit if there are exactly 0 photons in the idler using a long enough $\pi$-pulse resonant with the qubit frequency. The qubit is then measured and another $\pi$-pulse is applied at a frequency shifted by the dispersive shift $\chi$ such that the qubit state only changes if there is exactly one photon in the idler. The two qubit measurements lead to four possible outcomes: $gg$, $ge$, $eg$, and $ee$. If the measurement were perfect, only $gg$, $ge$, and $ee$ would be observed, corresponding to 2 or more, 1 and 0 photons respectively. In practice, the measurement is imperfect and to embrace these non-idealities, we depart from the proposal~\cite{Guha2009} and treat $\nu$ as an effective photon number which is set to one out of four values $\nu_{gg}$, $\nu_{ge}$, $\nu_{eg}$, and $\nu_{ee}$ conditioned on the qubit measurement outcomes.  The goal being to use $\nu$ as an estimator of the target state and not of the actual photon number, these four values are not limited to integer numbers and can be chosen at will. 
We repeat the experiment $M$ times to gather some statistics about $\nu$. Owing to the central limit theorem, the error exponent is then~\cite{Guha2009}
\begin{equation}
    \label{eq:errorexp}
    \mathcal{E} = \frac{\left(\expval{\nu^{(\mathrm{yes})}}-
                                       \expval{\nu^{(\mathrm{no})}}\right)^2}
                                     {2 (\sigma(\nu^{(\mathrm{yes})}) +  
                                         \sigma( \nu^{(\mathrm{no})}))^2},
\end{equation}

with $\expval{\nu^{(\mathrm{yes/no})}}$ and $\sigma(\nu^{(\mathrm{yes/no})})$ the average effective photon number and its standard deviation when the target is present or absent. 
For each value of the signal $N_S$ and noise $N_N$, we numerically fine tune the values $\nu_m$ in order to maximize the error exponent.

\paragraph*{Tuning up the quantum radar}

\begin{figure}[!hb]
    \centering
    \includegraphics[width=85mm]{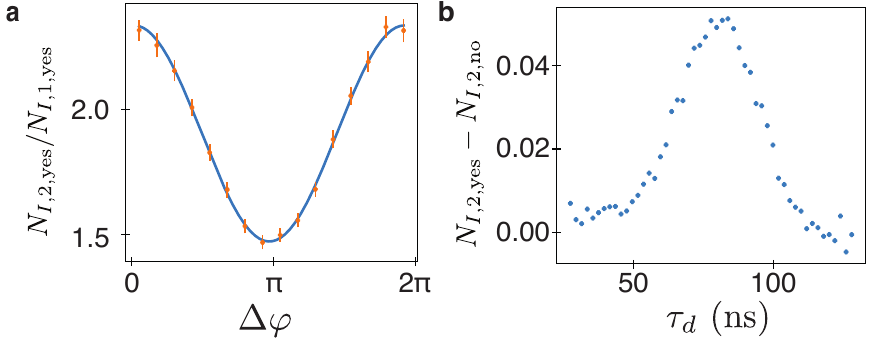}
    \caption{\label{fig:figure_2}Tuning up the interferometer. \textbf{a}) Dots: measured average photon number in the idler resonator after the second pump pulse $N_{I,2,\mathrm{yes}}$ as a function of the phase difference $\Delta\varphi$ when the target is present, with $N_S\approx 0.1$, and without thermal noise $N_N = 0$. The number is normalized by the measured average number of photons in the idler mode after the first pump pulse $N_{I,1,\mathrm{yes}}$. Line: cosine function fitted to the oscillation. \textbf{b}) Dots: measured average change of photon number in the idler resonator after the second pump pulse between present or absent target, and as a function of the delay $\tau_d$.}
\end{figure}

The exploitation of quantum correlations between signal and idler also requires finely tuning the pump pulse that recombines these modes. In contrast to the pump amplitude, the delay $\tau_d$ and phase offset $\varphi$ between the pump pulses (see \cref{fig:figure_1}c) can be chosen by operating the radar without added noise ($N_N = 0$), and at the largest signal setting ($N_S\approx 0.1$). With the target present, we measure the average number of photons in the idler mode after the first $(N_{I,1,\mathrm{yes}})$ and the second pulse $(N_{I,2,\mathrm{yes}})$ (see Methods). \cref{fig:figure_2}a shows the cosine dependence of the ratio $N_{I,2,\mathrm{yes}}/N_{I,1,\mathrm{yes}}$ as a function of the phase difference $\Delta\varphi=\varphi-\varphi_\mathrm{0}$ between the two-mode-squeezing operations for a delay $\tau_d=\SI{86}{ns}$. The phase $\varphi_0=-1.898$, corresponding to the maximal signal, depends on the electrical delay of the target and detuning of the pump. For the quantum radar experiment, we operate at $\Delta\varphi=0$. 
The cosine dependence originates from an interference. In fact, our experiment implements a new kind of $\mathrm{SU}(1,1)$ interferometer~\cite{Yurke1986, Flurin2012, Ou2020}, where one of the arms that host the TMSV is a stationary mode. In this particular case, the asymmetric loss probability $\kappa$ on one arm prohibits witnessing any remaining entanglement. 
We optimize $\tau_d$ at $\Delta\varphi=0$ by measuring how many extra photons are in the idler resonator after the second pulse when the target changes from absent to present. This idler population increase $N_{I,2,\mathrm{yes}}-N_{I,2,\mathrm{no}}$ is maximum for $\tau_{d, \mathrm{opt}} \approx \SI{86}{ns}$, see (\cref{fig:figure_2}b) which corresponds to the propagation delay of the signal to and back from the target.  

The joint measurement can be further optimized by tuning the amplitude of the second pump, which can be recast as a gain $G$ of the second two-mode squeezing operation. 
 
\begin{figure}
    \centering
    \includegraphics[width=85mm]{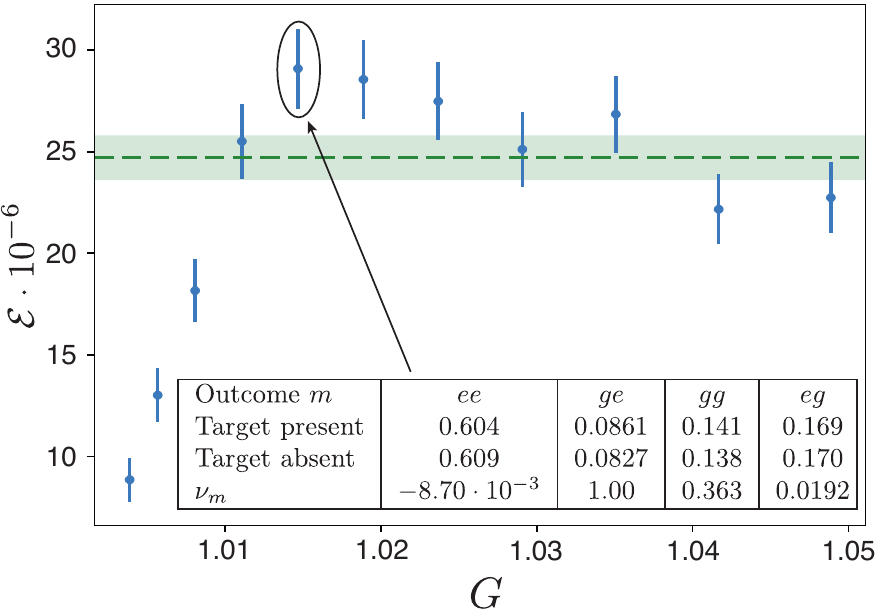} 
    \caption{\label{fig:figure_3}Observation of a quantum advantage for a microwave radar. Dots: measured error exponent $\mathcal{E}$ of the quantum radar as a function of the gain $G$ of the two-mode squeezing recombination of the signal and idler. Here, the number of signal and noise photons are independently measured to be $N_S = 3.53(4) \cdot 10^{-2}$ and $N_N = 10.8(3)$. Each point is obtained using 15 series of $5\cdot 10^5$ tries. After each series, $N_N$ and $N_S$ are re-calibrated. Green dashed line: quantum Chernoff bound providing the upper bound on the error exponent of any classical radar under the same conditions. The error bars and the colored area represent the uncertainties (see Methods). Inset: raw measurements for the highlighted point. For each possible outcome $m$, the table shows the fraction of occurrences where $m$ is found with the target being present or not, as well as the four values of $\nu$ that are used in \cref{eq:errorexp} to reach the highest error exponent. At this point, the quantum advantage is $Q=1.2(1)$. }
\end{figure}

An expression for the optimal gain $G$ is known for a given set of $N_S$, $N_N$ and $\kappa$~(see~\cite{Shi2022} and Methods), but we choose to empirically tune the gain $G$ to compensate for the non-idealities of our setup. We set $N_S$ and $N_N$ to particular values and measure the error exponent $\mathcal{E}$ for several values of $G$. For the settings of \cref{fig:figure_3}, it reaches a maximum $\mathcal{E}=2.9(2)\cdot 10^{-5}$ for a gain of about $G\approx 1.015$, which is close to the prediction by~\cite{Shi2022} of $G\simeq 1.016$.  

\paragraph*{Quantum advantage and inherent limitations}
In order to compute the quantum advantage $Q=\mathcal{E}/\mathcal{E}_\mathrm{cl}$, we now need to carefully calibrate the three parameters that set $\mathcal{E}_\mathrm{cl}$: the signal photon number $N_S$, the injected noise photon number $N_N$ and target reflectivity $\kappa$. Each parameter is determined during the same experimental run, using a dedicated protocol.

The signal photon number is set by the first squeezing operation, in which the circuit acts as a phase-preserving amplifier of gain $G_0$, giving $N_S = G_0 \nth{S} + (G_0-1) \left(1 + \nth{I} \right)$  with $\nth{S}$ and $\nth{I}$ the initial thermal populations of the signal and idler resonators. The pump amplitude is chosen to set the gain to small values $G_0\gtrsim 1$ to ensure $N_S\ll1$. To characterize $N_S$, we make use of the fact that the number of photons in the idler after the first squeezing operation is $N_I = N_S - \nth{S}+\nth{I}$. Determining $N_S$ thus only requires calibrating the initial thermal population of both signal and idler and measuring $N_I$ (see Methods). The thermal equilibrium population of the idler is about $1.5 \cdot 10^{-2}$. 
We further improve the purity of the TMSV state by initiating all the experimental realizations by sideband cooling the idler down to $\nth{I}=2.5(5) \cdot 10^{-3}$,  corresponding to a temperature of $29(1)~\mathrm{mK}$. We also measure an upper bound on $\nth{S}$ of $5 \cdot 10^{-3}$, which contributes to the error bars in Fig.~\ref{fig:figure_3}. 

To characterize $N_N$, we use the fact that when pumped at $\omega_\Delta = \omega_S - \omega_I$ with a large enough amplitude, the JRM induces a beam-splitter interaction between the idler and signal resonators which equilibrates the thermal fluctuations of the two modes. We can thus use the qubit to perform a steady-state measurement of the thermal population in the idler when noise and beam-splitter pump are injected to obtain $N_N$ (see Methods).

To precisely measure the target reflectivity $\kappa$, we use the superconducting device as a quantum vector network analyzer at the signal frequency. We send a coherent state via the auxiliary input line on the signal resonator that is either directly captured into the idler mode~\cite{Peronnin2019,dassonneville2020,Dassonneville2021} by using a pump at $\omega_\Delta$ or captured only after it reflected off the signal resonator, went through the target and came back into the resonator. The reflectivity is given by the ratio of the average amplitudes of the states captured into the idler resonator, which we characterize by performing a Wigner tomography of the idler mode (see Methods). We find $\kappa=3.02(8) \cdot 10^{-2}$ when the target is present and $3.2(9) \cdot 10^{-4}$ when absent.

In Fig.~\ref{fig:figure_3}, the measurements of $N_S=3.53(4)\cdot 10^{-2}$, $N_N=10.8(3)$ and $\kappa=3.02(8) \cdot 10^{-2}$ enable us to compute the upper bound on the error exponent that can be reached using coherent illumination: $\mathcal{E}_\mathrm{cl}=2.1(1)\cdot10^{-5}$.
This quantum radar thus beats the best possible classical one by a factor $Q=1.2(1)$, on par with what was achieved in optics~\cite{Zhang2015}. Note that taking into account the non-zero reflectivity when the target is absent would only lead to a slightly better quantum advantage since $\mathcal{E}_{\mathrm{cl}}$ would decrease by about 1\

\begin{figure}
    \centering
    \includegraphics[width=85mm]{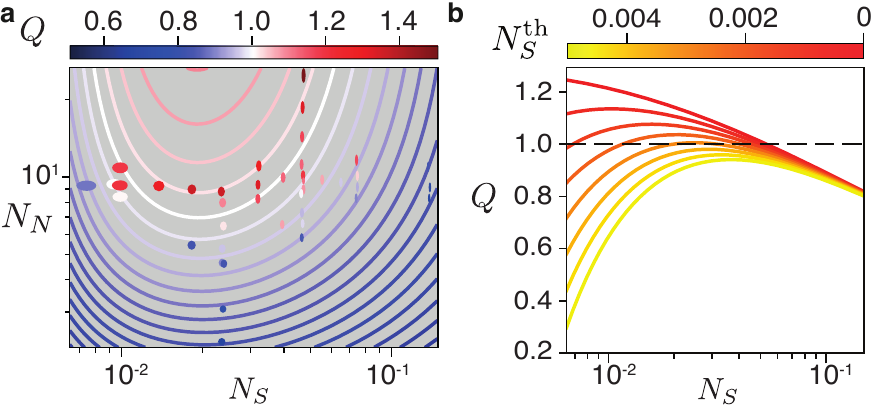}
    \caption{\label{fig:figure_4}Quantum advantage sensitivity to parameters. \textbf{a}) Contour plot of the predicted quantum advantage $Q$ as a function of the signal and noise photon numbers $N_S$ and $N_N$. The model (see Methods) is an extended version of Ref.~\cite{Guha2009} with a simplified model of photocounting. The superimposed colored dots represent the measured quantum advantage $Q$ as a function of signal and noise photon numbers $N_S$ and $N_N$. For each dot, we have measured the quantum advantage $Q$ as a function of receiver gain $G$ and only show its maximum value. The uncertainty on $Q$ is not shown but ranges from $4 \cdot 10^{-2}$ to $0.2$. The dots' width and height represent the $4\sigma$ uncertainties on $N_S$ and $N_N$. The initial thermal population in the signal $\nth{S}$ is set to $2 \cdot 10^{-3}$. \textbf{b}) Predicted quantum advantage as a function of the signal photon number $N_S$ for a fixed value of noise $N_N = 10$ for various initial signal thermal populations ranging from $0$ to $5 \cdot 10^{-3}$.}
\end{figure}

The quantum advantage we observe is obtained for a small signal photon number $N_S$ and a large noise photon number $N_N$. In order to determine the domain in the $N_S$, $N_N$ parameter space where a quantum advantage can be observed, we reproduce this measurement for various values of $N_S$ and $N_N$ and identify the maximal quantum advantage $Q$ as a function of receiver gain $G$, with the results shown in \cref{fig:figure_4}a. As these measurements and their associated calibrations take at least a few hours per point, we explore only a subset of the parameter space. Besides, the error exponent $\mathcal{E}_\mathrm{cl}=\kappa N_S/N_N$ gets smaller and smaller as $N_N$ increases or $N_S$ decreases so that it requires a longer measurement time.

From this measurement it appears that the quantum advantage increases with $N_N$ as expected. Guha and Erkmen~\cite{Guha2009} also predict that $Q$ increases at low $N_S$ until reaching its maximum values of $Q=2$. In our experiment we rather observe that $Q$ diminishes when $N_S$ becomes too small. 

We find that this behavior originates from the nonzero initial thermal populations $\nth{S}$ and $\nth{I}$ of the signal and idler modes respectively~\cite{DiCandia2021}. A model (see Methods) taking $\nth{S}$ and $\nth{I}$ into account and using an idealized version of our photocounting measurement is shown in \cref{fig:figure_4}a and qualitatively reproduces the experimental results in \cref{fig:figure_4}a. However, we note that the model systematically underestimates the measured quantum advantage. While the origin of this discrepancy remains an open question, the modeling of the measurement of the effective photon number $\nu$ could be a likely culprit. Note that for this figure, we set $\nth{S}$ to be $2 \cdot 10^{-3}$ which qualitatively reproduces our result better than the most pessimistic value of $5 \cdot 10^{-3}$ used in \cref{fig:figure_3} to demonstrate a quantum advantage. In \cref{fig:figure_4}b, we evaluate this model for different values of $\nth{S}$, and reveal how the window of signal photon number that exhibit a quantum advantage $Q>1$ shrinks, then disappears as $\nth{S}$ increases.

We thus find that this thermal population is a major limitation in our experiment, contrary to idler loss~\cite{Barzanjeh2015}. In our case, the latter only lowers the error exponent by $1-e^{-\gamma_I \tau_d}\approx 2\%$.

\paragraph*{Conclusion}
We have demonstrated an advantage of quantum radar versus classical radar in the microwave domain. The experiment reveals the crucial importance of the purity of the TMSV state used to illuminate the target. Beyond the loss of idler photons, this limitation imposes a stringent upper bound on the idler temperature. The experiment makes clear that using this quantum advantage in practical settings is a tremendous challenge. For instance, strategies that perform non-adaptative separate -- where the observables are determined before the experiment -- measurements of signal and idler at room temperature and use post-processing to extract correlations between the two~\cite{Barzanjeh2020,Hosseiny2022} cannot show a quantum advantage $Q>1$~\cite{Bradshaw2017,Shi2022}. 
Our work shows how superconducting circuits can provide quantum enhanced sensing in radar. 
While this exact scenario of quantum radar has limited applications~\cite{Shapiro2020, Sorelli2020,Jonsson2020,Jonsson2021}, it paves the way to demonstrations of other protocols measuring the range~\cite{Zhuang2022} or velocity of a target~\cite{Reichert2022}. Besides, our joint measurement could be replaced by a measurement of the signal followed by a feedforward to the idler~\cite{Shi2022}, which gives hope for an open air version of the quantum enhanced radar with a room temperature target. Another route consists in realizing a memory for many idler modes, using superconducting cavities~\cite{Chakram2021} or spin ensembles~\cite{Julsgaard2013}, in order to go beyond Q=2~\cite{Zhuang2017,Shi2022}. Using quantum correlations for enhanced sensing can also be applied to other research. For dark matter search it would be interesting to apply our demonstration to axion detection~\cite{Brady2022}.  For quantum communications, the quantum radar can be recast as the signaling of a bit of information (target present or not) through a noisy communication channel beyond the classical Shannon limit~\cite{Bennett2002, Hao2021, Shi2020}. Finally, the origin of a quantum advantage without residual entanglement is still a fascinating puzzle worth exploring further~\cite{Weedbrook2016,Bradshaw2017,Jo2021,Yung2020}.

\subsection*{Acknowledgments}
This work is part of Quantum Flagship project QMICS that has received funding from the European Union's Horizon 2020 research and innovation program under grant agreement No 820505.
We acknowledge IARPA and Lincoln Labs for providing a Josephson Traveling-Wave Parametric Amplifier. The devices were fabricated in the cleanrooms of ENS de Lyon, Collège de France, ENS Paris, CEA Saclay, and Observatoire de Paris. We thank Mikel Sanz, Mateo Casariego, Joonas Govenius, Jeff Shapiro, Pierre Rouchon and Daniel Est\`eve for fruitful discussions.

\subsection*{Author contributions}
R.A. performed the experiment and analysed the data. R.D. provided additional support for the experiment and analysis. T.P. fabricated the superconducting circuit and R.A. fabricated the target. R.A., R.D., A.B. and B.H. designed the experiment. B.H. supervised the project. All authors wrote the manuscript.

\begin{widetext}

\section*{Methods}

\subsection{Measurement setup and samples}
The cryogenic microwave setup is shown in \cref{fig:schematic_fridge}. The superconducting device in the Cryoperm shield is the exact same device that was used in the experiments of Refs.~\cite{Peronnin2019,dassonneville2020,Dassonneville2021}. The 12~m delay line is made of two microwave cables in series. They are made of a \SI{3.58}{mm} semi-rigid coaxial cable constructed with silver plated copper clad steel inner conductor, solid PTFE dielectric and tinned aluminum outer conductor. At the output of the signal resonator port, and right beyond the directional coupler that injects the pump and noise into the signal resonator port on demand, a tee with two bandpass filters routes the probe signal at $\omega_S\approx 10.2~\mathrm{GHz}$ towards the target while it routes the reflected pump at $\omega_P\approx 14~\mathrm{GHz}$ into a termination load.
The spying line was not used during the experiment.
\begin{figure}[h!]
    \centering
    \includegraphics[width=12cm]{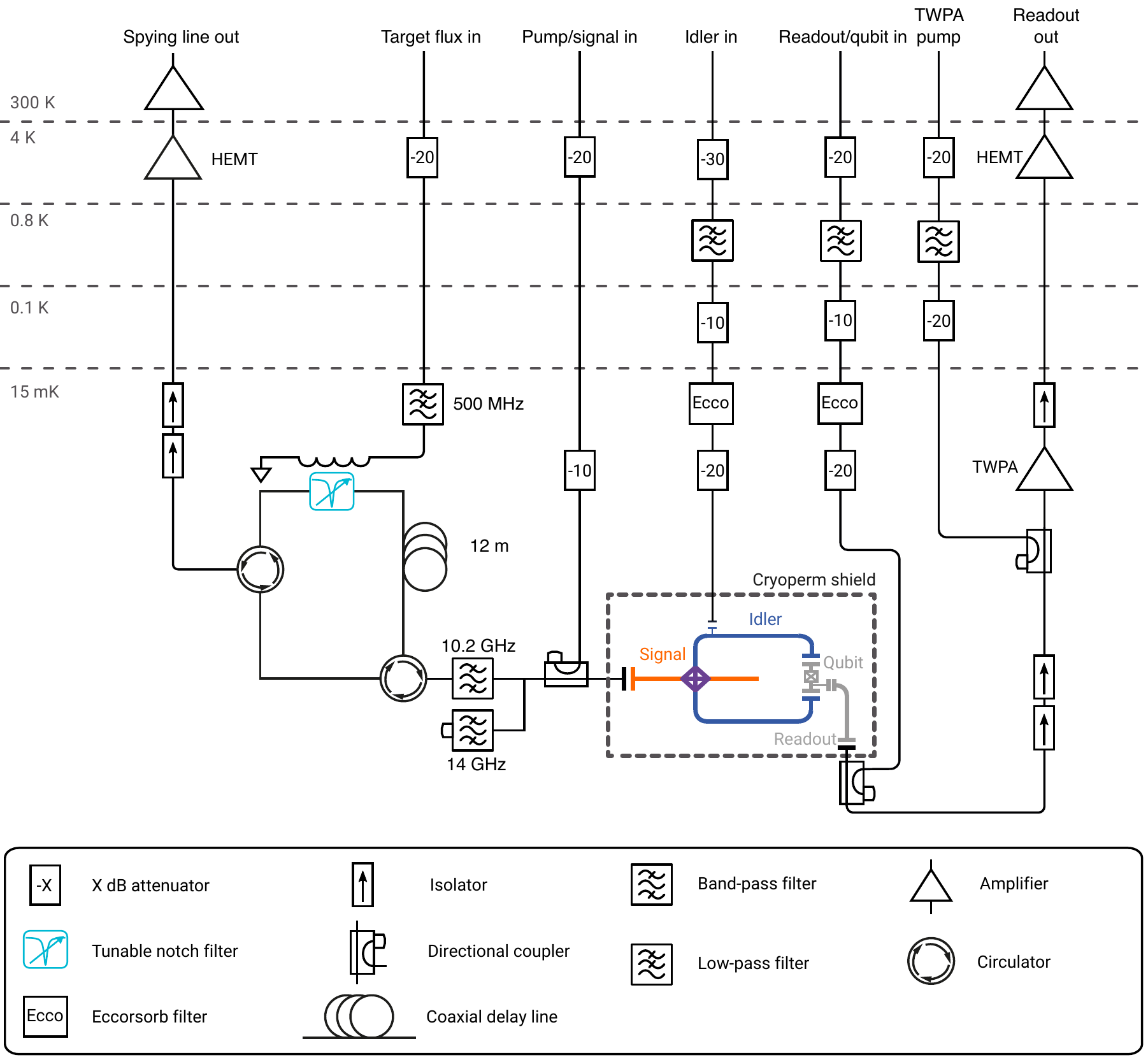}
    \caption{Schematic of the wiring inside the Bluefors LD250 dilution refrigerator used for the experiment with a base temperature at $15~\mathrm{mK}$. The Josephson Traveling Wave Parametric Amplifier (TWPA) was graciously provided by the Lincoln Lab. }
    \label{fig:schematic_fridge}
\end{figure}

A key component of the target is realized by a tunable notch filter. It is a stub filter made of a superconducting $\lambda/2$ resonator that shorts the transmission line to ground when on resonance with the signal (Fig.~\ref{fig:fluxMapTarget}a). The tunability comes from the two Josephson junctions in a loop (SQUID) that terminate the resonator. This device is made of sputtered Tantalum on a sapphire chip while the Josephson junctions and the loop are made using e-beam evaporated Al/AlOx/Al. A flux line is able to flux bias the loop fast enough so that one out of two measurements is performed in the present or absent configuration.

\begin{figure}[h!]
    \centering
    \includegraphics[width=12cm]{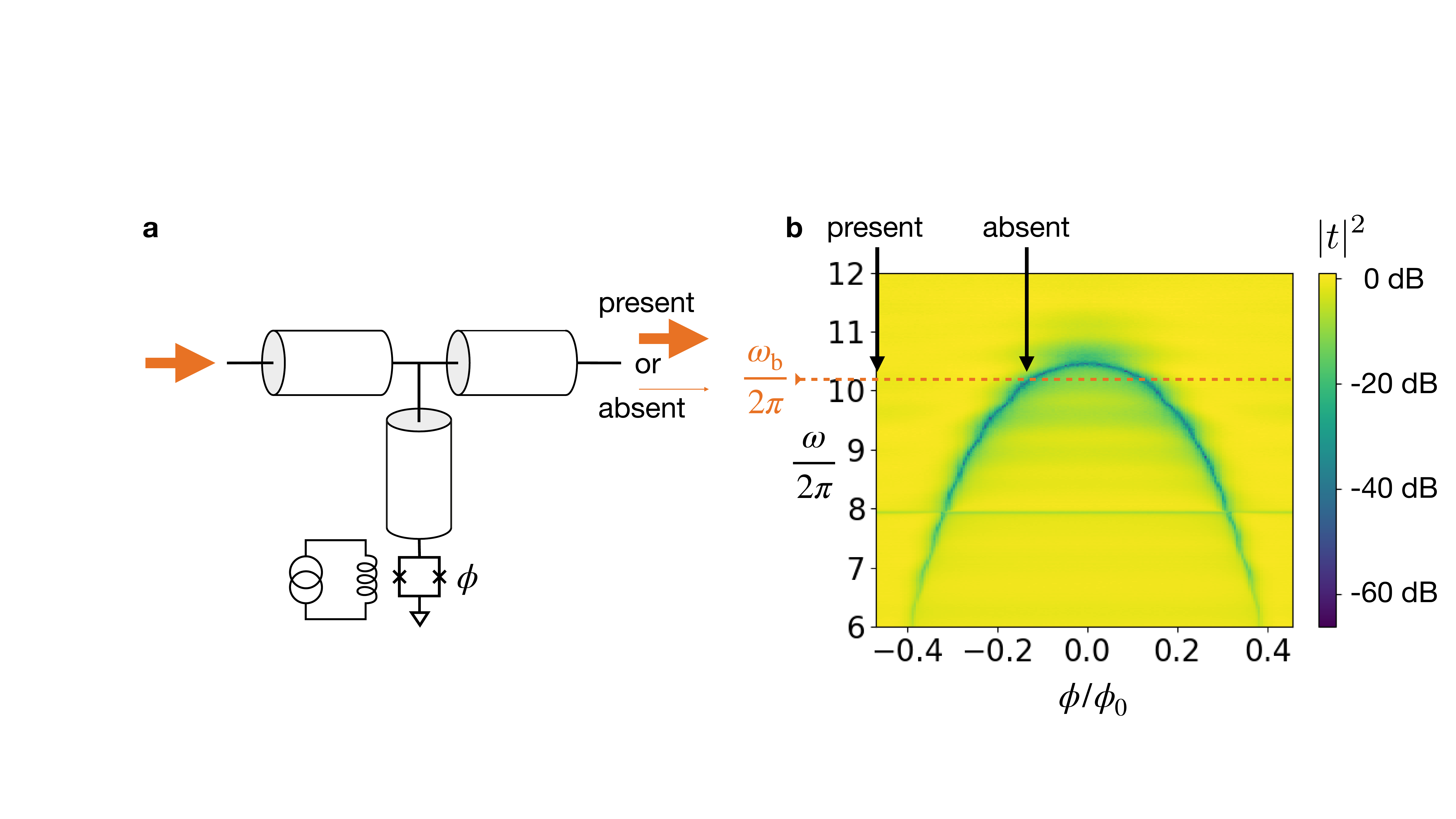}
    \caption{\label{fig:fluxMapTarget}\textbf{a}) Schematics of the tunable notch filter. \textbf{b}) Measured magnitude of the transmission coefficient $t$ across the filter as a function of flux bias $\phi$ and signal frequency $\omega$. The coefficient is normalized to a value measured out of resonance. A dashed line indicates the signal frequency.}
    
\end{figure}
Before being placed in the final setup described in \cref{fig:schematic_fridge}, the tunable filter part of the target was tested in a separate dilution refrigerator and its transmission was measured as a function of the current applied through the flux line. In Fig.~\ref{fig:fluxMapTarget}b is shown the measured transmission $|t|^2$ as a function of signal frequency and flux threading the loop. We measure a 3-dB rejection bandwidth of around \SI{100}{MHz}, an isolation of around \SI{20}{dB} and a tunability of the central frequency over several \si{GHz}.
In the experiment, circulators ensure that the signal that gets out of the signal resonator port first reaches the tunable notch filter and only comes back through the delay line if the target is present. 

\begin{figure}[h!]
    \centering
    \includegraphics[width=12cm]{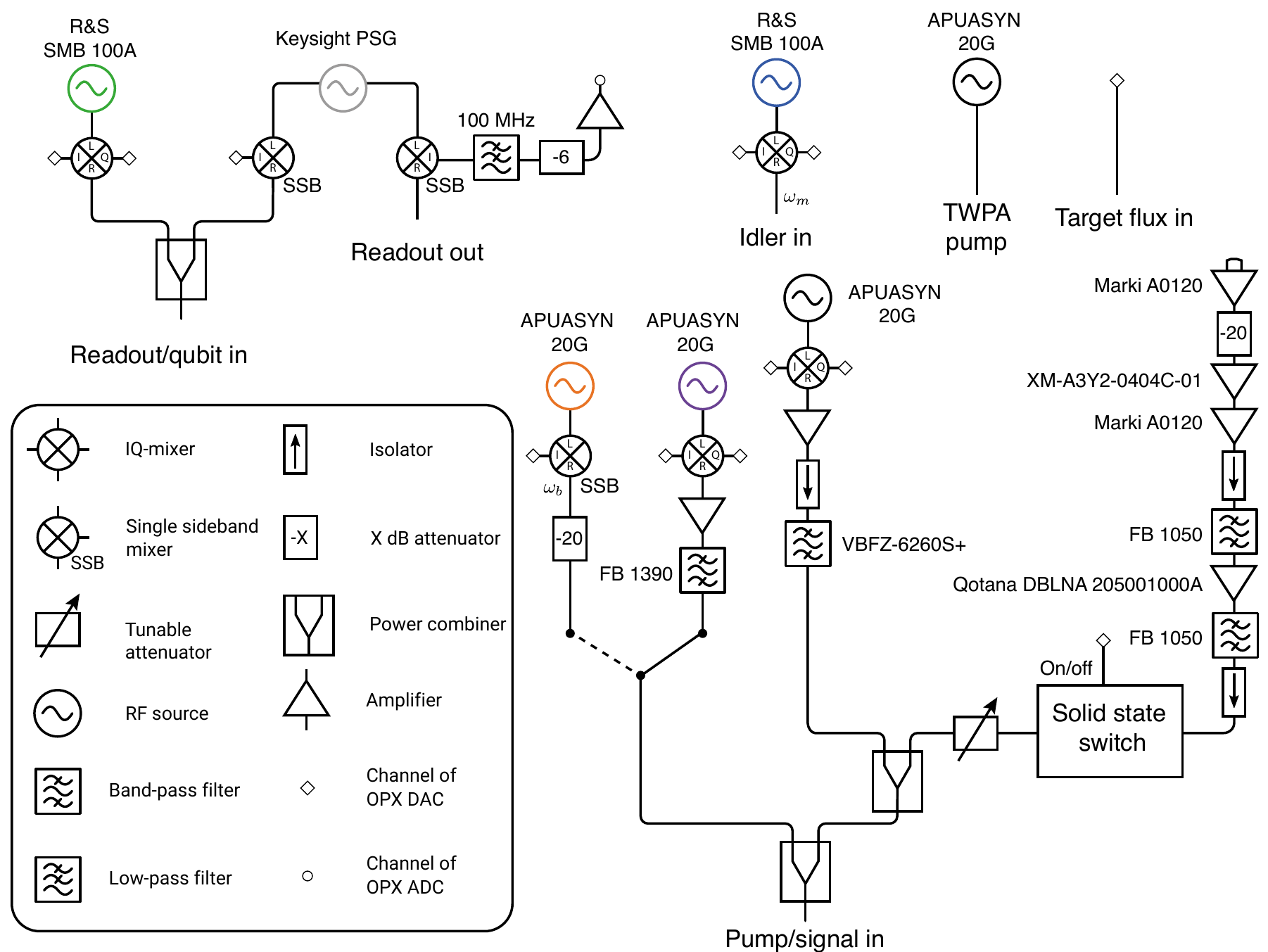}
    \caption{\label{fig:schematic_rt}Schematic of the control electronics. The experiment is controlled by an OPX from Quantum Machines.}
\end{figure}

\subsection{Idler mode photo-counting}

To probe the average photon number in the idler resonator, hence measuring $N_S$ or $N_I$ depending on context, we use the dispersive interaction $-\chi m^\dag m \ketbra{e}$ with strength $\chi/2\pi = \SI{4.75}{MHz}$ between the idler resonator and the transmon qubit, whose resonance frequency is \SI{4.23036}{GHz}. The qubit lifetime $T_1 = \SI{6.5}{\mu s}$ and coherence time $T_2 = \SI{12}{\mu s}$ put it in the photon-number resolved regime~\cite{Schuster2007}. In the experiment, we use three different ways to perform the photocounting of the idler mode. The first one is detailed in \cref{fig:figure_1}c. The others are detailed below.

To measure $N_I$ when it is not larger than 2, we use a technique based on Ramsey interferometry explained in Ref~\cite{dassonneville2020}. It consists in performing a $\pi/2$ pulse on the qubit, waiting a varying amount of time $t$, sending another $\pi/2$ pulse and finally measuring the qubit using homodyne detection $s_+(t)$ of the readout resonator output. In order to avoid experimental drifts in gain and phase, we interleave this measurement with another one where the second $\pi/2$ pulse is a $-\pi/2$ pulse, which gives a measurement record $s_-(t)$. We then compute $s(t) = s_+(t) - s_-(t)$. A typical measurement is shown in \cref{fig:popRamsey}. Because of the dispersive coupling between the idler resonator and the qubit, we observe oscillations of the readout record $s(t)$. Finally, we can fit those oscillations $s(t)$. Assuming that the idler is in a thermal state, and for $t \ll T_2$ or $N_I\ll 1$, we find
\begin{equation}
    s(t) \propto e^{-t/T_2} \sum_{k=0}^\infty \frac{N_I^k}{{(N_I+1)}^{k+1}} \cos(t (\chi k + \beta k^2)),
    \label{eq:popRamsey}
\end{equation}
with $\beta/(2\pi) = \SI{70}{kHz}$ the next higher order non-linear term $-\beta (m^\dag m)^2 \ketbra{e}$ in the Hamiltonian and $T_2 = \SI{12}{\micro s}$ the qubit decoherence time. The factors $\frac{N_I^k}{{(N_I+1)}^{k+1}}$ are the probability to find $k$ photons in a thermal state with average photon number $N_I$.  

When $N_I$ is larger, we use another method based on measuring the resonator relaxation towards its equilibrium population by monitoring the probability of having exactly 0 photons in the resonator.

To measure this probability,

we use a long $\pi$-pulse on the qubit that is selective on the presence of 0 photons in the cavity and then repeatedly measure the qubit population $P_e(t)$ for various waiting times $t$ between the initialization of the idler and the photon number selective $\pi$-pulse. By assuming that the idler is initially in a thermal state with average photon number $N_I$, we find that 
\begin{equation}
    P_e(t) =   \frac{1}{N_I e^{-t/T_1} + (1-e^{-t/T_1})\nth{I} + 1} \approx  \frac{1}{N_I e^{-t/T_1} + 1},
    \label{eq:p0relax}
\end{equation}
with $T_1 = \SI{4.1}{\micro s}$ the relaxation time of the idler mode. We can then fit the measured qubit excitation to this equation to find the average photon number initially contained in the idler mode.

\subsubsection{Calibration of \texorpdfstring{$\nth{I}$}{the idler thermal population}}

Using the Ramsey interferometry technique described above, we measured a thermal equilibrium population of $1.5(1) \cdot 10^{-2}$ for the idler mode, which corresponds to an approximate temperature of \SI{41}{mK}. To improve the performance of the radar, we actively cool down the idler using a beam-splitter interaction between the idler mode and a higher frequency mode activated by pumping at the difference of the two frequencies. Since this other mode has a much lower quality factor than the idler resonator and the beam-splitter interactions tends to even the number of photons, we are able to cool the cavity down to $\nth{I} = 2.5(5) \cdot 10^{-3}$. By chance, this cooling transition is merely \SI{79}{MHz} above of the two-mode squeezing transition enabling us to use the same mixer and lines for initial cooling and radar operation. All of the error exponent measurements we present are preceded by this \SI{1.2}{\micro \second} long cooling pulse.

\begin{figure}[h!]
    \centering
    \includegraphics[width=96mm]{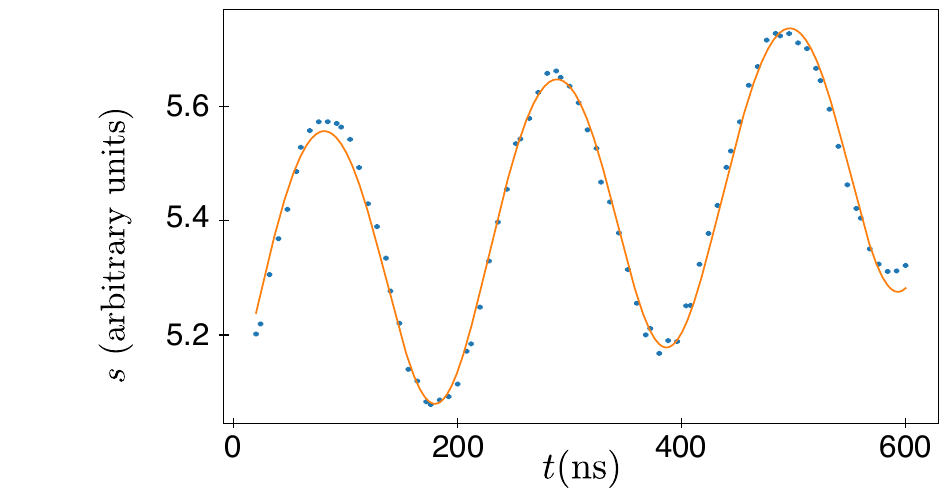}
    \caption{\label{fig:popRamsey}Typical result of an average signal measurement sequence using the Ramsey interferometry technique right after the first two-mode squeezing operation of the radar sequence. Dots: measured signal $s$. Solid line: fit of the oscillations using \cref{eq:popRamsey} with $N_I = 0.104(1)$.}
\end{figure}
\subsubsection{Calibration of \texorpdfstring{$N_S$}{Nₛ}\label{sec:signalCal}}

The calibration of the signal photon number $N_S$ is realized by measuring the average number of photons $N_I$ in the idler right after the first two-mode squeezing operation. The two are related by $N_I - \nth{I} = N_S - \nth{S}$. 

Finally, to convert this average number of idler photons $N_I$ into an average number of signal photons $N_S$ we need to know the difference between the thermal populations of signal and idler. While we were able to measure the thermal population of the idler with a relatively good precision to $\nth{I} = 2.5(5) \cdot 10^{-3}$, we were only able to place an upper bound of $5 \cdot 10^{-3}$ on the number of equilibrium thermal photons of the signal using a technique similar to the one described in \cref{sec:calib_nn}.

We also use this measurement to estimate the receiver gain $G$. Indeed, the number of photons $N_I$ we measure after a two-mode squeezing operation of gain $G$ is given by 
$N_I = G \nth{I} + (G-1) \left(1 + \nth{S} \right)$ leading to $G = (1+N_I+\nth{S} )/(1+\nth{I}+\nth{S})$.

\begin{figure}[h!]
    \centering
    \includegraphics[width=96mm]{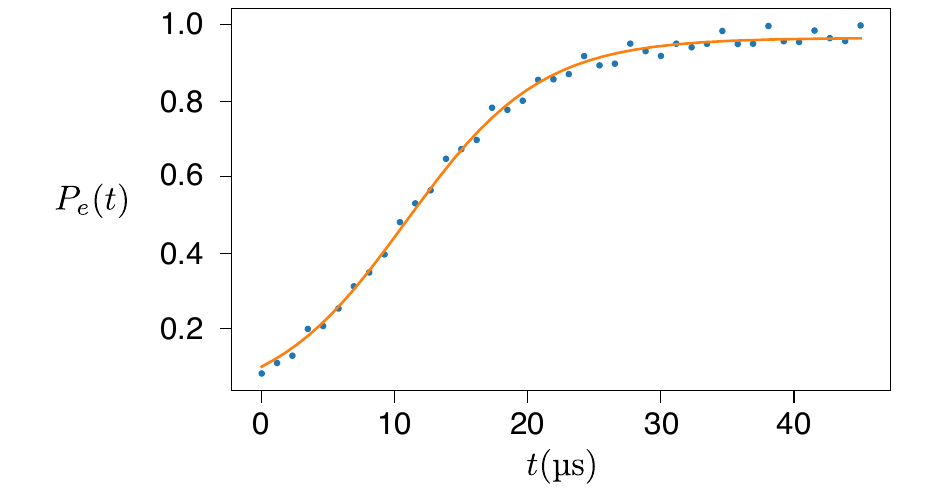}
    \caption{\label{fig:p0relax}Typical measurement of the noise photon number $N_N$ using the relaxation method. Dots: excited population of the qubit as a function of waiting time $t$ after the measurement sequence. Solid line: fit of the relaxation \cref{eq:p0relax} with $N_N = 8.6(5)$.}
\end{figure}
\subsubsection{Noise setup and calibration of \texorpdfstring{$N_N$}{Nₙ}\label{sec:calib_nn}}
As shown in \cref{fig:schematic_rt}, the noise photons $N_N$ are generated at room temperature by amplifying the Johnson-Nyquist noise of a \SI{50}{\ohm} resistor using a chain of amplifiers. To avoid saturating the final amplifiers or overloading the cooling capacity of the dilution refrigerator, bandpass filters are used to suppress the noise outside of the signal frequency window. The filters used are Marki FB 1050 with a \SI{1.5}{GHz} bandwidth which is much larger than the \SI{20}{MHz} bandwidth of the signal resonator making the noise perfectly thermal from the point of view of the signal. To adjust the noise, an electrically tunable attenuator is used as well as a solid-state switch (HMC-C019) which is able to turn the noise on after the generation of the signal/idler pair but before the signal possibly comes back from the target.

To calibrate the noise in-situ, while the noise is turned on, we first activate the beam-splitter interaction between idler and signal resonators by pumping the JRM at $\omega_\Delta = \omega_S - \omega_I$ which equalizes the photon number population inside both resonators. Once a steady state is reached, we switch off this pump and measure the average number of photons in the idler $N_N$ using the relaxation method described above.

\subsection{Calibration of \texorpdfstring{$\kappa$}{κ}\label{sec:calib_kappa}}

\begin{figure}[h]
    \centering
    \includegraphics{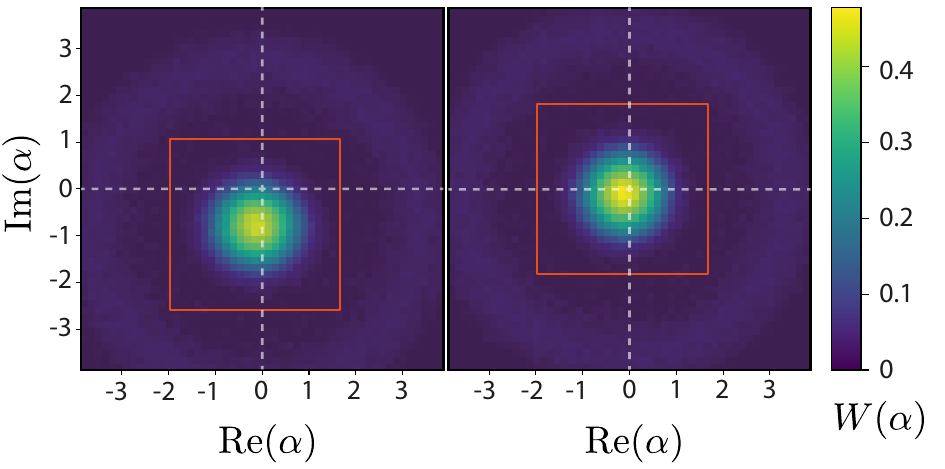}
    \caption{\label{fig:kappa_calib}Typical Wigner tomograms used in the calibration of the target reflectivity $\kappa$. The red square delimits the area used to find the average amplitude (see \cref{sec:calib_kappa}). Left: Measured Wigner function of the idler after a signal-idler swap performed right after a coherent drive of the signal resonator. Right: Measured Wigner function of the idler after a signal-idler swap performed after a reflection on the target when present.}
\end{figure}

To measure the target reflectivity $\kappa$, we implemented a sort of narrowband vector network analyzer (VNA) using the superconducting circuit as a sensor: first, we send a short coherent pulse (a \SI{28}{ns} long wavepacket shaped by a hyperbolic secant) through the directional coupler, the wavepacket then enters the signal resonator where we can choose to either measure it or let it bounce back, through the target and back to the signal resonator again with the attenuation $\kappa$ we want to estimate. To measure the average amplitudes of this wavepacket before and after going through the target, we swap the signal with the idler either before or after the wavepacket goes through the target  and then perform a full Wigner tomography of the idler state using the qubit. By taking the average amplitude weighted by the Wigner function, we can find the amplitude $\alpha_1$ and $\alpha_2$ of the incident and reflected signals and deduce the target reflectivity $\kappa = \abs{\frac{\alpha_2}{\alpha_1}}^2$. Note that the results are independent of the swap efficiency since the same swap sequence is used for the incident and reflected waves. We measure $\kappa_\mathrm{yes} = 3.02(8) \cdot 10^{-2}$ when the target is nominally present and a residual $\kappa_\mathrm{no} = 3.2(9) \cdot 10^{-4}$ when the target is nominally absent which agrees well with our independent \SI{20}{dB} isolation measurement.

The Wigner tomography is performed by first displacing the idler with an amplitude $-\alpha$ and then measuring its parity using the qubit. These parity measurements consist in applying a fast, unconditional, $\pi/2$ pulse on the qubit followed by a waiting time of $\pi/\chi$ and another $\pi/2$ pulse before reading out the state of the qubit. For additional robustness, we interleave sequences using a final $\pi/2$ pulse with sequences using a final $-\pi/2$ pulse as described in~\cite{Dassonneville2021}. A large faint ring with a radius of around $3$ appears in the Wigner tomograms (see \cref{fig:kappa_calib}). We attribute it to a breakdown of the dispersive approximation for such large excitations. Since we compute the average amplitude $\expval \alpha$ using $\expval{\alpha} = \int W(\alpha) \alpha \dd{\alpha}$, this ring would introduce a bias on the measurement. To eliminate this bias, we truncate the measured Wigner function to a smaller square roughly centered on the average amplitude. This smaller square is represented in red in \cref{fig:kappa_calib}.

\subsection{Quantum radar model}

When the initial state is generated by a squeezing operation with gain $G_0$, the signal and idler modes form a gaussian state with a zero mean described by the covariance matrix of the two pairs of creation and annihilation operators $(\hat a_S^\dag, \hat a_S)$ and $(\hat a_I^\dag, \hat a_I)$:  $V_E = \expval{{(\hat a_S^\dag ~ \hat a_I^\dag ~\hat a_S ~\hat a_I)}^\dag(\hat a_S^\dag ~ \hat a_I^\dag ~\hat a_S ~\hat a_I)}$

\begin{equation}
   V_E = \pmqty{N_S + 1 & 0 & 0 & N_C \\
    0 & N_I + 1 & N_C & 0 \\
    0 & N_C & N_S & 0 \\
    N_C & 0 & 0 & N_I}
\end{equation}

with $N_C = \sqrt{G_0(G_0-1)}( 1 + \nth m + \nth b)$, $N_S = G_0 \nth b + (G_0 - 1) (\nth m + 1)$ and $N_I = (G_0 - 1) (\nth b +1) + G_0 \nth m$.

The attenuation by the noisy target transforms the operator $\hat a_S$ into a reflected $\hat a_R = \sqrt{\kappa} \hat a_S + \sqrt{1 - \kappa} \hat a_N$ (when the target is absent, we take $\kappa = \kappa_\mathrm{no}$) with $\hat a_N$ the operator describing a thermal field with average photon number $\frac{N_N}{1-\kappa}$. Hence, at the receiver, the state is still gaussian with zero mean and its covariance matrix reads
$V_R = \expval{{(\hat a_R^\dag ~ \hat a_I^\dag ~\hat a_R ~\hat a_I)}^\dag(\hat a_R^\dag ~ \hat a_I^\dag ~\hat a_R ~\hat a_I)}$
\begin{equation}
    V_R = \pmqty{ \kappa N_S + N_N + 1 & 0 & 0 & \sqrt{\kappa} N_C \\
    0 & N_I + 1 & \sqrt{\kappa}N_C & 0 \\
    0 & \sqrt{\kappa}N_C & \kappa N_S + N_N & 0 \\
    \sqrt{\kappa}N_C & 0 & 0 & N_I}.
\end{equation}

Finally, after the recombination step between the reflected signal and idler with a two-mode squeezing operation of gain $G$, the state present in the idler mode is a gaussian state with zero mean and with an annihilation operator $\hat c = \sqrt{G} \hat a_I + \sqrt{G-1}\hat a_R^\dag$. Before being measured, the idler thus contains an average number of photons 

\begin{align*}
    \expval{c^\dag c}  &= G \expval{\hat a_I^\dag \hat a_I} + (G-1) \expval{\hat a_R \hat a_R ^\dag} + \sqrt{G(G-1)} \left(\expval{\hat a_I^\dag \hat a_R^\dag} + \expval{\hat a_R \hat a_I}\right)\\
    &= G N_I + (G-1) (1 + \kappa N_S + N_N) + 2 \sqrt{\kappa G (G-1)} N_C.
\end{align*}

The last term shows that the quantum correlations $N_C$ can at this point be accessed by measuring the average number of photons in the idler mode. This is the key point enabling a quantum advantage in quantum radar.
Using \cref{eq:errorexp}, and the fact that the $\hat c$ mode has thermal statistics we can compute the error exponent assuming an ideal photo-counting measurement of $\hat c$:

\begin{equation}
    \mathcal E = \frac{{\left(\expval{c^\dag c}_\mathrm{yes} - \expval{c^\dag c}_\mathrm{no}\right)}^2}{2 {\left(\sqrt{\expval{c^\dag c}_\mathrm{yes}^2 + \expval{c^\dag c}_\mathrm{yes}} + \sqrt{\expval{c^\dag c}_\mathrm{no}^2 + \expval{c^\dag c}_\mathrm{no}}\right)}^2}
\end{equation}
with $\expval{c^\dag c}_\mathrm{yes} = G N_I + (G-1) (1 + \kappa N_S + N_N) + 2 \sqrt{\kappa G (G-1)} N_C$ and $\expval{c^\dag c}_\mathrm{no} = G N_I + (G-1) (1 + N_N)$.

\subsection{Choice of receiver gain \texorpdfstring{$G$}{G}}
\begin{figure}[h!]
    \centering
    \includegraphics[width=6cm]{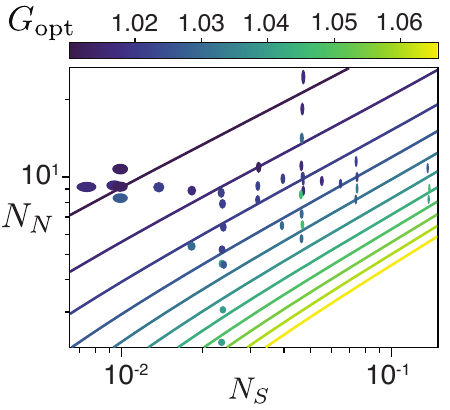}
    \caption{\label{fig:best_gain} Optimal recombination gain as a function of signal and noise photon numbers $N_S$ and $N_N$. The contour plot is the theoretical expression \cref{eq:opt_gain} for the optimal gain $G_\mathrm{opt}$. The superimposed colored dots represent the gain that was used in the experiment to get the best error exponent using a coarse optimization. The dots' width and height represent the $4\sigma$ uncertainties on $N_S$ and $N_N$.}
\end{figure}

The receiver used in this experiment is often called an optical parametric amplifier (OPA) in the literature. It works by recombining the reflected signal and the idler using a two-mode squeezing operation.

The gain $G_\mathrm{opt}$ associated to this squeezing that maximizes the Fisher information of the receiver has been calculated in Ref~\cite{Shi2022}. The optimal gain is $G_\mathrm{opt} = \max(1, G^*)$ with

\begin{equation}
    \label{eq:opt_gain}
    G^* = 1 + \frac{\sqrt{N_S(N_S+1)(N_N + \kappa N_S)(N_N + \kappa N_S + 1)} + N_S(N_S+1)}{(N_N + (\kappa-1)N_S)(N_N + (\kappa + 1)N_S + 1)}
\end{equation}

In practice, contrary to the assumptions made by the authors of~\cite{Shi2022}, the emitted TMSV is impure and the final photon counting is imperfect. Hence, the best gain was empirically chosen for each pair of signal photon number $N_S$, and noise photon number $N_N$. Given the long measurement time required, only a few values of gains were tried for each point. With this coarse optimization, the values of gain $G_\mathrm{opt}$ leading to the best error exponent are shown in \cref{fig:best_gain} on top of the theoretical predictions. The gain value is a rough estimate as it was not as frequently and precisely calibrated as the main parameters $N_S, N_N$ and $\kappa$. Nevertheless, the optimal gain measured does behave as predicted and decreases when either the signal or the noise increases.

\subsection{Uncertainties calculation}
The uncertainty on the estimation of the quantum advantage $Q = \frac{\mathcal E}{\mathcal{E}_\mathrm{cl}}$ comes from two sources: the uncertainties on the estimation of the quantum Chernoff bound for the classical error exponent $\mathcal{E}_\mathrm{cl}$ and those coming from the measurement of the error exponent $\mathcal{E}$ of the quantum radar.

Let us examine the estimation of the uncertainty of the quantum Chernoff bound $\mathcal{E}_\mathrm{cl}$ first. The three parameters used in the computation are the target reflectivity $\kappa$, the number of photons in the signal beam $N_S$ and the number of noise photons $N_N$. The measurements of those three parameters are described in \cref{sec:calib_kappa,sec:signalCal,sec:calib_nn}. The uncertainties we used on $N_S$ and $N_N$ come from the non-linear fitting routine which means that they are mostly statistical and do not take into account imperfections in the measurement protocols. To estimate the uncertainty on $\kappa$, different probe amplitudes are used and the distribution of results allows us to make sure that $\kappa$ is independent of power (at least in the low power range we are considering) as well as provide a statistical uncertainty on the value of $\kappa$. These three uncertainties are then propagated through the expression of $\mathcal{E}_\mathrm{cl} = \frac{\kappa N_S}{4 N_N}$ by assuming no correlations between the three quantities which gives a total uncertainty of 

\begin{equation}
    \Delta \mathcal{E}_\mathrm{cl} = \frac{\kappa N_S}{4 N_N} \sqrt{ \frac{\Delta^2(\kappa)}{\kappa^2} + \frac{\Delta^2(N_S)}{N_S^2} + \frac{\Delta^2(N_N)}{N_N^2} }
\end{equation}

for $\mathcal{E}_\mathrm{cl}$.

For the measured error exponent $\mathcal E$ of the quantum radar, the uncertainties on the measured mean and variance of the effective photon number $\nu$ are required. Assuming a large number of repetitions $M$ and a Gaussian distribution for $\nu$ (which is true if $M$ is large enough to fulfill the central limit theorem conditions), the uncertainties on $\expval{ \nu}$ and $\sigma(\nu)$ are given by $\Delta \left(\expval{\nu} \right) = \sigma(\nu) / \sqrt M$ and $\Delta(\sigma(\nu)) = \sigma(\nu) / \sqrt{2M}$. 
To propagate these uncertainties through \cref{eq:errorexp} to the uncertainty on $\mathcal E$, we use the two following rules
\begin{align*}
    \Delta \left(\frac{X}{Y}\right) = \abs{\frac{X}{Y}}\sqrt{{\left(\frac{\Delta X}{X}\right)}^2 + {\left(\frac{\Delta X}{X}\right)}^2 + 2\frac{r(X, Y) \Delta(X) \Delta(Y)}{X Y}} \\
    \Delta \left(X+Y\right) = \sqrt{\Delta(X)^2 + \Delta(Y)^2 + 2r(X, Y) \Delta(X) \Delta(Y)}
\end{align*}
where $r(X,Y) = \frac{\mathrm{Cov}(X,Y)}{\Delta(X) \Delta Y}$ is the Pearson correlation coefficient between the stochastic variables $X$ and $Y$. We therefore need to estimate three Pearson coefficients in the experiment:
\begin{enumerate}
    \item  $r\left[\left(\expval{\nu^{(\mathrm{yes})}}-\expval{\nu^{(\mathrm{no})}}\right)^2,
                                       (\sigma(\nu^{(\mathrm{yes})}) + \sigma(\nu^{(\mathrm{no})}))^2\right]$ between the signal and the noise of the quantum radar. One might expect some positive correlations between the two since both should be affected similarly by technical drifts in the parameters, and one should be proportional to the other. Since estimating those correlations experimentally is quite difficult, we choose to consider the worst case scenario that maximizes the uncertainty on $\mathcal E$ and use the value $0$ in the calculations.
    \item  $r\left[ \expval{\nu^{(\mathrm{yes})}}, \expval{\nu^{(\mathrm{no})}} \right]$ between the measured signals when the target is present and absent. One might expect some systematic biases (such as the value of $\kappa$ or $N_S$ drifting) to push this coefficient above $0$ but we take again the most pessimistic value, which is $0$ once again.
    \item  $r\left[ \sigma(\nu^{(\mathrm{yes})}), \sigma(\nu^{(\mathrm{no})}) \right]$ between the noise level when the target is present and absent. Just like the previous case, one might also expect some positive correlation between the two distributions. Here the most pessimistic assumption is to take this coefficient to be $1$ and assume full correlation.
\end{enumerate}

These three assumptions allow us to put a reasonable upper bound on the value of the uncertainty of $\mathcal E$

\begin{equation}
    \Delta(\mathcal E)  = \frac{1}{\sqrt M} \left(\frac{\expval{\nu^{(\mathrm{yes})}}-
                                       \expval{\nu^{(\mathrm{no})}}}
                                     {\sigma(\nu^{(\mathrm{yes})}) +  
                                         \sigma(\nu^{(\mathrm{no})})}\right)^2 \sqrt{ \frac{1}{2} + \frac{\sigma(\nu^{(\mathrm{yes})})^2 +  
                                         \sigma(\nu^{(\mathrm{no})})^2}{\left(\expval{\nu^{(\mathrm{yes})}}-
                                       \expval{\nu^{(\mathrm{no})}}\right)^2} }.
\end{equation}
\end{widetext}
\bibliography{biblio_clean.bib}
\end{document}